\documentclass[10pt,twocolumn]{wlscirep}

\usepackage[utf8]{inputenc}
\usepackage[T1]{fontenc}
\usepackage[switch]{lineno}
\usepackage[english]{babel}
\usepackage[euler]{textgreek}
\usepackage[version=4]{mhchem}
\usepackage{amsmath}
\usepackage{hyperref}
\usepackage{graphicx}
\usepackage{color}
\usepackage{amssymb}
\usepackage{mathtools}
\usepackage{multicol}
\usepackage{physics}
\usepackage{tikz}

\usetikzlibrary{fadings}
\usetikzlibrary{decorations.pathmorphing,patterns}
\usetikzlibrary{shapes.misc}
\usetikzlibrary{shapes.misc}
\usepackage{soul}

\newcommand{\onlinecite}[1]{\hspace{-1 ex} \nocite{#1}\citenum{#1}}

\newlength{\thelinewidth}
\thelinewidth=\textwidth
\title{Magnon-driven dynamics of a hybrid system excited with ultrafast optical pulses}

\author[1,2,*]{N. Crescini}
\author[2,3]{C. Braggio}
\author[2,3]{G. Carugno}
\author[3,4]{R. Di Vora}
\author[1]{A. Ortolan}
\author[1]{G. Ruoso}

\affil[1]{INFN - Laboratori Nazionali di Legnaro, Viale dell'Universit\`a 2,  Legnaro (PD) 35020, Italy}
\affil[2]{DFA - Università degli Studi di Padova, Via Marzolo 8, Padova 35131, Italy}
\affil[3]{INFN - Sezione di Padova, Via Marzolo 8, Padova 35131, Italy}
\affil[4]{DSFTA - Università di Siena, Via Roma 56, Siena, 53100, Italy}
\affil[*]{nicolo.crescini@phd.unipd.it}


\begin{abstract}
The potential of photon-magnon hybrid systems as building blocks for quantum information science has been widely demonstrated, and it is still the focus of much research. We leverage the strengths of this unique heterogeneous physical system in the field of precision physics beyond the standard model, where the sensitivity to the so-called “invisibles” is currently being boosted by quantum technologies.
Here, we demonstrate that quanta of spin waves, driven by tiniest, effective magnetic field, can be detected in a large frequency band using a hybrid system as transducer.
This result can be applied to the search of cosmological signals related, for example, to cold Dark Matter, which may directly interact with magnons.
Our model of the transducer is based on a second-quantisation two-oscillators hybrid system, it matches the observations, and can be easily extended to thoroughly describe future large-scale ferromagnetic haloscopes. 
\end{abstract}

\begin{document}

\flushbottom
\maketitle

\thispagestyle{empty}

\section*{Introduction}
\begin{figure*}[h!]
	\centering
\scalebox{0.7}
{
	\begin{tikzpicture}
	\node at(5,6) { \includegraphics[width=\textwidth]{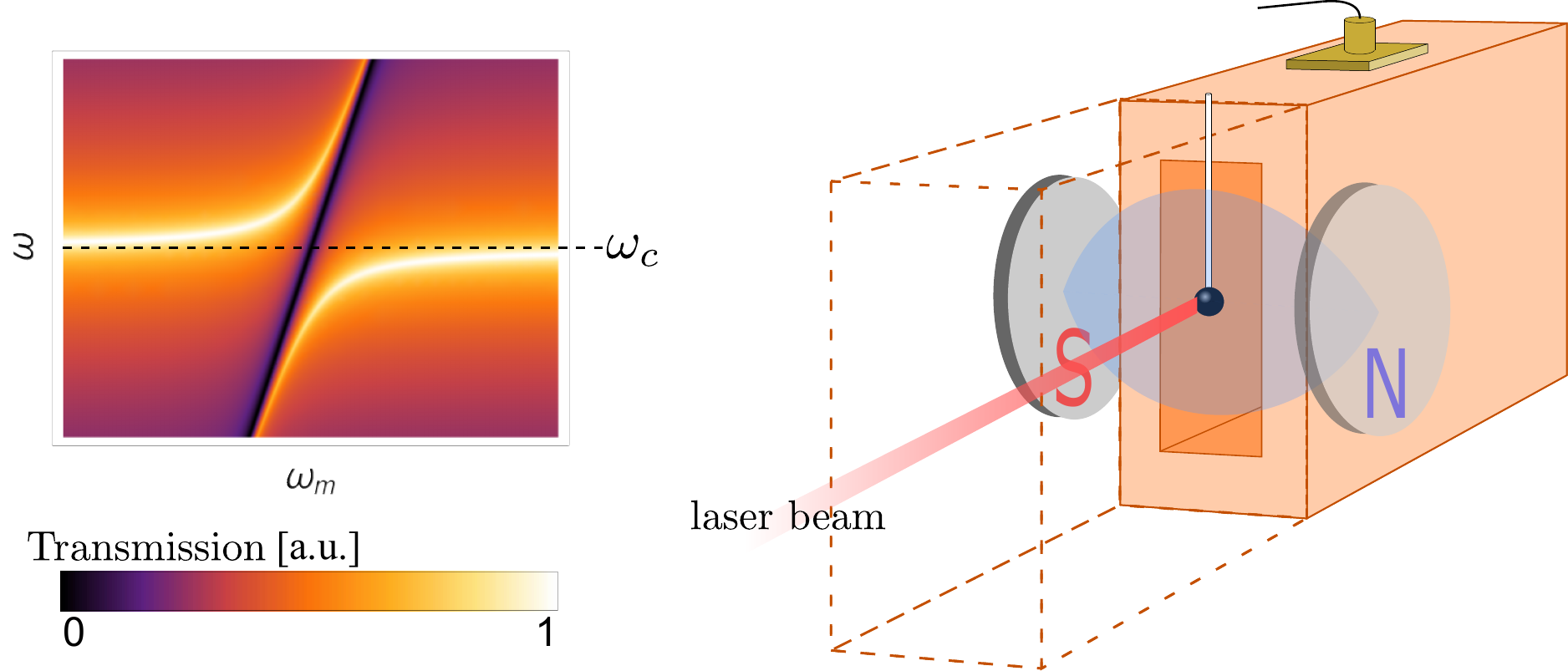} };
	\draw	[draw=teal, fill=teal!20]	(0.5,0) circle [radius=.5];
	\draw	[<->, draw=teal, thick]	(-.5,-.5) to [in=-150,out=-30] (1.5,-.5);
	\node at (0.5,0) {\Large \textcolor{teal}{a}};
	\node [below] at (0.5,-1) {\Large \textcolor{teal}{$\omega_a$}};
	\draw	[draw=black!30!green, fill=green!20] 	(6,0) circle [radius=1];
	\draw	[<->, draw=black!30!green, thick]	(5,-1) to [in=-150,out=-30] (7,-1);
	\node at (6,0) {\Large \textcolor{black!30!green}{m}};
	\node [below] at (6,-1.5) {\Large \textcolor{black!30!green}{$\omega_m$}};
	\draw	[draw=orange, fill=orange!20]	(10,0) circle [radius=1];
	\draw	[<->, draw=orange, thick]	(9,-1) to [in=-150,out=-30] (11,-1);
	\node at (10,0) {\Large \textcolor{orange}{c}};
	\node [below] at (10,-1.5) {\Large \textcolor{orange}{$\omega_c$}};
	\draw (1,0) to (2,0);
	\draw[decoration={aspect=0.3, segment length=1.5mm, amplitude=1.5mm,coil},decorate,thick] (2,0) to (4,0);
	\draw (4,0) to (5,0);
	\node [above, align=left] at (3,0.1) {\Large weak\\ \Large coupling};
	\draw[decoration={aspect=0.3, segment length=3mm, amplitude=3mm,coil},decorate,thick] (7,0) to (9,0); 
	\node [above, align=left] at (8,.5) {\Large strong\\ \Large coupling};
	\node at(-4,9) {\Large \bf (a)};
	\node at(4.5,9) {\Large \bf (b)};
	\node at(-1,1) {\Large \bf (c)};
	\end{tikzpicture}
}
	\caption{{\bf (a)} Avoided level crossing of a photon-magnon hybrid system. The curve is typically obtained by recording the cavity transmission spectrum at several values of the static magnetic field, which determines the ferrimagnetic resonance $\omega_m$ according to $\omega_m=\gamma B_0\simeq(2\pi\times28\,\mathrm{GHz/T})\times B_0$, with $\gamma$ gyromagnetic ratio of the electron. 
	{\bf (b)} Scheme of the apparatus used in this work. The magnetic material is hosted inside a microwave cavity, and polarised with the field generated by an electromagnet. A laser pulse excites the magnetostatic modes of a YIG sphere, including the Kittel mode, allowing for investigating the HS dynamic response to a direct excitation of its magnetic component.  {\bf (c)} Coupled oscillators representation of the ferromagnetic haloscope, in which strong coupling is realised between an rf cavity mode (frequency $\omega_c$) with the Kittel mode ($\omega_m$) to reach cosmologically relevant sensitivity. The axion-DM field acts as a tiniest classical rf field on the HS, and it is thus represented by a weakly coupled oscillator.}
	\label{fig:ho}
	\end{figure*}
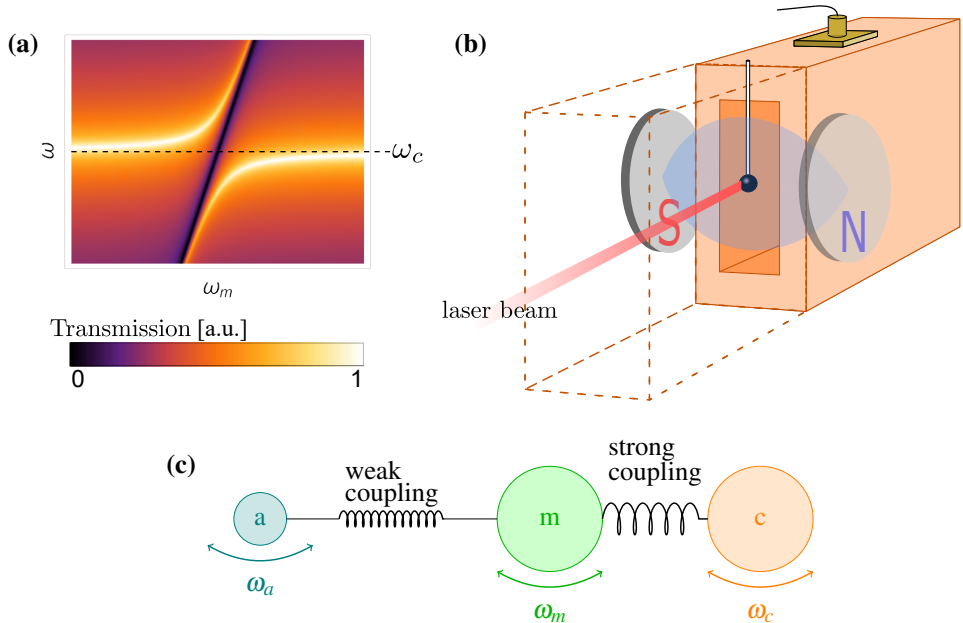

In the last decades, precision magnetometry emerged as a promising probe of physics beyond the standard model\cite{doi:10.1146/annurev.nucl.012809.104433,Kirch2018}. Amongst all, GHz frequency magnetometers can probe spin-related effects through the use of the electron spin resonance techniques.
Recent advances in the field are due to quantum computing research\cite{nakamura_review}, which allowed to reach the sensitivity to detect single quanta of magnetization in macroscopic samples using photon-magnon hybrid systems (HSs)\cite{single_magnons,Lachance-Quirion425}. In these devices, the coherent interaction between photons and magnons is increased to such an extent that they can no longer be considered as separate entities, and the HS is described within the cavity quantum electrodynamics framework. Photons are confined within a microwave cavity hosting a magnetized sample as shown in Fig.\,\ref{fig:ho}b, whose Larmor frequency $\omega_m$ is adjusted close to the one of a suitable cavity mode $\omega_c$. 
In the strong coupling regime, in which the coupling strength is much larger than the related linewidths, photons and magnons are in a hybrid magnon-polariton state induced by magnon Rabi oscillation. Such a cooperative spin dynamics is governed by the physics of coupled harmonic oscillators with beating periods much shorter than dissipation times\cite{PhysRevLett.105.140501}. 
As under these conditions the two oscillators exchange energy, the dispersion plot of the system displays the quantum phenomenon of avoided crossing: when $\omega_c\simeq\omega_m$, instead of two intersecting lines, the system dispersion relation exhibits an anticrossing curve\cite{landau1981quantum,Kittel2004} as shown in Fig.\,\ref{fig:ho}a.

These HSs have been realised and extensively studied in the last decade\cite{Clerk2020,PhysRevLett.111.127003,PhysRevLett.113.083603,PhysRevLett.113.156401,PhysRevApplied.2.054002,Zhang2015}, and applied to the development of quantum memories\cite{Khabat,Kurizki3866}, microwave to optical photon conversion\cite{PhysRevLett.118.107205,PhysRevB.93.174427,PhysRevA.92.062313,PhysRevLett.113.203601,Kimble2008}, detection of single magnons\cite{Lachance-Quirion425}, or in spintronics\cite{Chumak2015}.
Furthermore, the investigation of non-Hermitian quantum mechanics\cite{moiseyev2011non} is currently pursued with photon-magnon HSs.
Here we focus on a HS devised to probe weak and persistent effective rf fields, which is the physics case of dark matter (DM) axions\cite{BARBIERI2017135}.

The axion is an hypothetical particle introduced as a consequence of the strong CP problem solution found by Peccei and Quinn \cite{pq,weinberg1978new,wilczek1978problem}. It is a light pseudo-Goldstone boson arising from the breaking of the Peccei-Quinn symmetry at extremely high energies $f_a\simeq10^{12}\,$GeV. Since its mass and couplings are proportional to $1/f_a$ the axion is extremely light and weakly interacting \cite{DINE1983137,PhysRevLett.43.103,SHIFMAN1980493,DINE1981199}. Relevant quantities of them may have been produced in the early universe, qualifying the axion as a viable candidate of cold DM that would account for the whole density $\rho_\mathrm{DM}=0.45$\,GeV/cm$^3$ of the Milky Way’s halo\cite{spergel170wilkinson, abbott1983cosmological, turner1990windows, raffelt1996stars, PRESKILL1983127}. 
A suitable mass range for the axion is $m_a=10^{-(3\divisionsymbol5)}\,$eV\cite{bonati, berkowitz2015lattice, borsanyi2016calculation, petreczky2016topological,diCortona2016}, hence  its de Broglie wavelength is of the order of meters and allows a coherent interaction with macroscopic systems. Together with the high occupation number $\rho_\mathrm{DM}/m_a$, this allows to treat the axion DM field as a classical rf field.
The presence of axions can be tested with earth-based precision measurements \cite{PhysRevLett.51.1415,redondo,axion_searches} probing observables sensitive to the presence of DM axions, i.\,e. the rf-power in a microwave cavity under a static magnetic field\cite{PhysRevLett.118.061302,refId0,MCALLISTER201767,PhysRevLett.104.041301,caldwell2017dielectric,PhysRevLett.120.151301,PhysRevLett.124.101303} or the magnetization of a sample\cite{BARBIERI2017135, PhysRevX.4.021030, quaxepjc, tobar,Garcon:2017ihx}; we focus on this last case.
The described HSs emerge as a natural choice to detect the axion effective field as the variation of a sample magnetization, and an apparatus measuring the power deposited in a HS by DM axions, and hereafter called $P_{ac}$, is therefore called ferromagnetic haloscope.
As the underlying interaction is tiniest, the ferromagnetic haloscope relies on maximising signal power $P_{ac}$, which translates to hosting in the limited volume of GHz-frequency cavities a sample with the highest possible electron spin density. In addition, the hybrid modes $Q$-factor should not be much lower than the axion figure of merit ($2\times10^6$). By using yttrium iron garnet (YIG) as magnetic sample, these two requirements are well satisfied, owing to its high spin density ($2\times10^{28}$\,spins/m$^3$) and small damping constant ($\sim 10^{-5}$).

In this work we theoretically analyse the response of a photon-magnon HS to a direct excitation of the magnetic component using a second-quantisation model, and experimentally test the model results with an apparatus involving an optical excitation of the magnetic material.
We obtain the first experimental demonstration of the broad frequency tunability of the ferromagnetic haloscope, a crucial property that previously had only been theoretically analysed\cite{tobar}. This  precision detector at the low energy frontier of particle physics can probe the axion coupling with electron spins in mass ranges of few $\mu$eV, as large as that probed by Primakoff haloscopes\cite{PhysRevLett.118.061302,doi:10.1063/1.4938164,PhysRevLett.120.151301,PhysRevLett.124.101303}, with the advantage of a much simpler frequency scan system. The latter is, in this case, accomplished by varying the amplitude of the external magnetic field, instead of moving tuning rods in ultra-cryogenic environments.
While Primakoff haloscopes are based o the coupling of axions to photons, ferromagnetic haloscopes search for axions through their electron interaction. Since testing both these couplings is a way to distinguish between different axions, the two detectors provide a complementary insight into the different models.

In addition, the knowledge we gain on the dynamic response of this photon-magnon HS, allows for the optimising the spin magnetometer, as we can maximise the collected axion power with different magnetising fields.

\section*{Results}
\subsection*{Theoretical model}
To understand the dynamics of a ferromagnetic haloscope and optimize its operation, it is useful to model it. 
The system parameters are the resonant frequencies of the cavity mode $\omega_c$, the one of the Kittel mode $\omega_m$, and their linewidths $\gamma_c$ and $\gamma_m$, respectively.
We use a second quantisation formalism\cite{walls2007quantum,scully1999quantum,dutra2005cavity,walther2006cavity,doi:10.1119/1.4792696} in natural units to write the Hamiltonian of this system as the one of two coupled oscillators, photon and magnon, and an external axion field $a$ that interacts only with the magnon. In the rotating-wave approximation it results
	\begin{align}
	\begin{split}
	&H= \omega_m m^+m+ \omega_c c^+c +g_{cm}(mc^+ + m^+c) \\
	&+ g_{am}\sqrt{N_a}(me^{-i\omega_a t}+m^+e^{i\omega_at}),
	\label{eq:ham}
	\end{split}
	\end{align}
where $m$ ($m^+$), $c$ ($c^+$) are the destruction (creation) operators of magnons and photons respectively, and $g_{cm}$ and $g_{am}$ are the magnon-photon and axion-magnon couplings. 
A scheme of the considered system is shown in Fig.\,\ref{fig:ho}c.
The last term in Eq.\,(\ref{eq:ham}) represent the interaction of the axion with the magnon, $N_a$ is the axion number $N_a = \langle a^+a\rangle$ and its mass fixes the frequency $\omega_a$. We are assuming $N_a>>1$ and treating axions as a classical external field perturbing the photon-magnon system. 
The effect of the axion field is to deposit energy in the form of spin-flips (i.\,e. magnons), which, with the rate $g_{cm}$, are converted into photons that can be collected as rf power. 

Starting from the Hamiltonian in Eq.\,(\ref{eq:ham}), the evolution of the system can be derived with the Heisenberg-Langevin equations\cite{walls2007quantum,scully1999quantum} for the mean values of magnon $\langle m\rangle$ 
	\begin{equation}
	i \dot{\langle m \rangle} = (\omega_m-i\gamma_m/2) + g_{cm}c + g_{am}\sqrt{N_a}e^{-i\omega_at}
	\label{eq:hl_m}
\end{equation}
and cavity photon $\langle c\rangle$  state
	\begin{equation}
	i \dot{ \langle c \rangle} = (\omega_c-i\gamma_c/2) + g_{cm}m,
	\label{eq:hl_p}
	\end{equation}
where $\gamma_{c,m}$ accounts for dissipations to the thermal reservoir. 
Given the high occupation number of the magnon and photon states, we solve the equations in a semiclassical approach, and neglect quantum correlations.
Eq.s\,(\ref{eq:hl_m}) and (\ref{eq:hl_p}) can be recast as a matrix differential equation 
	\begin{equation}
	i\dot{M}= \Big[
	\begin{pmatrix}
	\omega_m-\frac{i\gamma_m}{2}	&	g_{cm}				\\
	g_{cm}					&	\omega_c-\frac{i\gamma_c}{2}	\\
	\end{pmatrix}
	+ 
	g_{am} \sqrt{N_a}e^{-i\omega_a t}
	\begin{pmatrix}
	1 \\
	0
	\end{pmatrix}
	\label{eq:ham_matrix} \Big] M,
	\end{equation}
where $M$ is the two-component vector $M=( \langle m\rangle, \langle c \rangle )$.
If we rewrite $M$ as a plane wave $M=A\exp(-i\omega_at)$, Eq.\,(\ref{eq:ham_matrix}) can be recast as
	\begin{align}
	\begin{split}
	& \Big[
	\begin{pmatrix}
	\omega_a 	&	0			\\
		0		&	\omega_a 	
	\end{pmatrix}
	-
	\begin{pmatrix}
	\omega_m -\frac{i\gamma_m}{2}	&	g_{cm}				\\
	g_{cm}					&	\omega_c-\frac{i\gamma_c}{2}	
	\end{pmatrix}
	\Big] = KA  \\
	&= g_{am}\sqrt{N_a}
	\begin{pmatrix}
	1	\\
	0	\\
	\end{pmatrix}
	,
	\end{split}
	\end{align}
yielding the amplitude of $M$
	\begin{equation}
	A=g_{am}\sqrt{N_a}K^{-1}
	\begin{pmatrix}
	1	\\
	0	
	\end{pmatrix}
	\end{equation}
from which it is possible to extract the (2,1) component describing the coupling between the axion field and the cavity
	\begin{align}
	\begin{split}
	A_{21}&=g_{am}\sqrt{N_a}(K^{-1})_{21}\\
	&=\frac{g_{cm}g_{am}\sqrt{N_a}}{(\omega_a-\omega_m+i\gamma_m/2)(\omega_a-\omega_c+i\gamma_c/2)-g_{cm}^2}.
	\end{split}
	\end{align}
The observable of our apparatus is the power deposited in the resonant cavity by the axion field, which can be calculated through the quadrature operator $z \equiv (c^++c)/ \sqrt{2  \omega_c}$, and results
	\begin{align}
	\begin{split}
	& P_{ac}=\frac{\gamma_c}{2}\langle\dot{z}^2\rangle = \frac{\gamma_c \omega_a^2}{4\omega_c}|A_{21}|^2 \\
	&= \frac{\gamma_c \omega_a^2}{8\omega_c} \frac{g_{cm}^2 g_{am}^2  N_a}{|(\omega_a-\omega_m+\frac{i\gamma_m}{2})(\omega_a-\omega_c+\frac{i\gamma_c}{2})-(\frac{g_{cm}}{2})^2|^2}.
	\label{eq:pac}
	\end{split}
	\end{align}
The obtained expression of $P_{ac}$ fully describes the dynamics of the system and can be used to maximise the ferromagnetic haloscope sensitivity.
It is interesting to consider the case of the power deposited by an axion field on resonance with one of the two hybrid modes of the system at frequencies
	\begin{equation}
	\omega_\pm = \frac{\omega_c+\omega_m}{2} \pm \sqrt{\Big(\frac{\omega_c-\omega_m}{2}\Big)^2 +\Big( \frac{g_{cm}}{2}\Big)^2}.
	\label{eq:omegapm}
	\end{equation}
Since the system collects power at the hybrid modes frequencies, to infer the bandwidth of the apparatus one needs to recast $\omega_m$ in terms of $\omega_\pm$ by inverting Eq.\,(\ref{eq:omegapm}).
Substituting the expression into Eq.\,(\ref{eq:pac}), $P_{ac}$ results
	\begin{align}
	\begin{split}
	&P_{ac}(\omega_\pm)\big\rvert_{\omega_a=\omega_\pm}=\\
	&\frac{\gamma _c\omega_\pm^2 g_{cm}^2 g_{am}^2 N_{a}/8\omega _c}{ \left| \left(\omega_\pm+\frac{i \gamma _c}{2}-\omega _c\right) \left(\omega_\pm+\frac{i \gamma _m}{2}-\frac{\omega_\pm^2-\omega _c \omega_\pm-(g_{cm}/2)^2}{\omega_\pm-\omega _c}\right)-\big(\frac{g_{cm}}{2}\big)^2\right|^2}.
	\label{eq:p_bw}
	\end{split}
	\end{align}

\begin{figure}[h!]
\centering
\begin{tikzpicture}
\node	at(0,10.5) {\includegraphics[width=.45\textwidth]{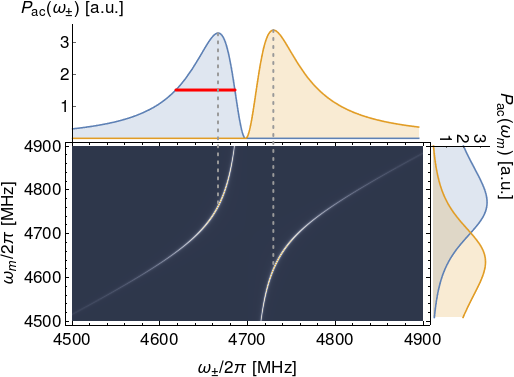}};
	\draw[black,thick,rounded corners,fill=gray!10] (-4,4.5) rectangle (4,7);
	\node[right] at(-3.75,6.5) {$\omega_c=(2\pi)\,4.7\,$GHz};
	\node[right] at(-3.75,6) {$\gamma_c=(2\pi)\,1.1\,$MHz};
	\node[right] at(-3.755,5.5) {$\gamma_m=(2\pi)\,3.5\,$MHz};
	\node[right] at(-3.75,5) {$g_{cm}=(2\pi)\,26.5\,$MHz};
	\node[right] at(0.25,5.5) {\textcolor{blue}{$g_{am}=(2\pi)\,10^{-13}\,$MHz}};
	\node[right] at(0.25,5) {\textcolor{blue}{$N_a=(2\pi)\,10^{24}$}};
	\node[above right,text width=2,align=center] at(0.25,5.75) {\textcolor{red}{Dynamical \\bandwidth}};
	\node[above right] at(2,6) {\textcolor{red}{$\simeq 64\,$MHz}};
	
	\node at(-4,13.5) {\bf (a)};
	\node at(-4,7.25) {\bf (b)};
\end{tikzpicture}
\caption{Deposited axion power as a function of the HS normal modes $(\omega_\pm)$. (a) The main figure is the anticrossing plot $P_{ac}(\omega_\pm,\omega_m)$ given by Eq.\,(\ref{eq:pac}), while the upper and right plots are the projection of the deposited power as function of $\omega_\pm$ and $\omega_m$, respectively. The upper plot, representing $P_{ac}(\omega_\pm)$, shows the apparatus bandwidth in blue for $\omega_-$ and in orange for $\omega_+$. The bandwidth is calculated as the FWHM of the $\omega_-$ curve of $P_{ac}(\omega_\pm)$, and its experimentally measured value is reported in red. (b) In the gray box, the experimental parameters used for the calculation are reported in black, $g_{am}$ and $N_a$ (values in blue) have been arbitrarily chosen, and the dynamical bandwidth is shown in red.}
\label{fig:bw}
\end{figure}
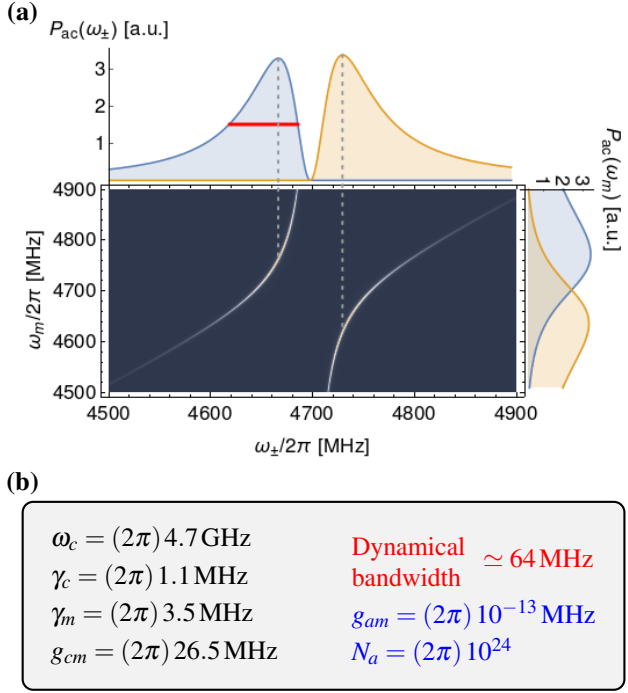

The deposited power is plotted in Fig.\,\ref{fig:bw} for the parameters of our experimental apparatus, and is in agreement with previously reported results\cite{tobar,Flower_2019}.
The haloscope collects power on two separated axion-mass intervals, each being one order of magnitude broader than the resonance linewidth, demonstrating a wideband tunability of the haloscope.
Hereafter, we call dynamical bandwidth the frequency interval that can be scanned by changing the Larmor frequency, highlighted in red in Fig.\,\ref{fig:bw} for the lower frequency hybrid mode.

The parameters used for this calculation are arbitrarily chosen as they do not influence the experimental results, but we mention that they are related to the axion model and to some cosmological parameters. In particular, the axion number $N_a$ depends on the axion number density and on the volume of magnetic material; under the assumption that DM is entirely composed of axions, at the considered frequency we have $\rho_\mathrm{DM}/m_a\simeq 2\times10^{13}\,\mathrm{axions/cm^3}$. The axions figure of merit depends on their thermal distribution and, as they are cold DM, the energy dispersion is smaller than its mean by a factor $2\times10^6$.  The coupling $g_{am}$ is proportional to the axion-electron coupling constant $g_{aee} \simeq 3\times10^{-11}(m_a/1\,\mathrm{eV})$, which slightly depends on the considered axion model, and is a number inversely proportional to the axion mass. For more details on the axion-to-magnon conversion scheme see Ref.s\,\onlinecite{BARBIERI1989357,BARBIERI2017135,quaxepjc}.

\subsection*{Experimental validation}
	\begin{figure*}[h!]
	\centering
		\begin{tikzpicture}
		\node at(-4,-5) { \includegraphics[width=.45\textwidth]{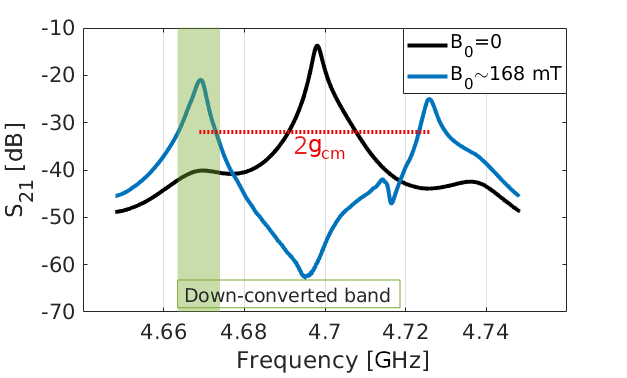} };
		\node at(4,-5) { \includegraphics[width=.45\textwidth]{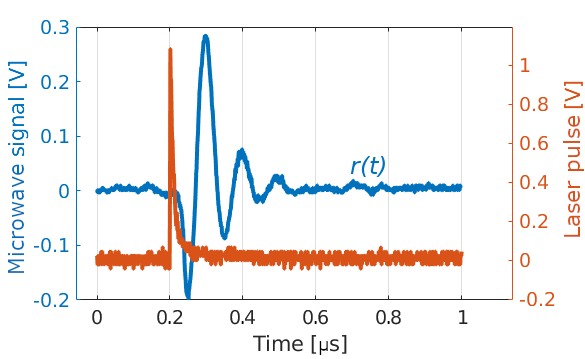} };
		\node at(-8,-3) {\large \bf (a)};
		\node at(0,-3) {\large \bf (b)};
		\end{tikzpicture}
	\caption{{\bf (a)} Transmission spectrum of the cavity ($B_0=0$) and of the hybrid system ($B_0=168\,$mT). The frequency band where the signal is down-converted is evidenced in green. The frequency separation between the hybrid modes is related to the coupling strength between cavity photons and magnons. {\bf (b)} Oscilloscope traces of the relevant signals. The laser pulse acts as acquisition trigger (orange) for the down-converted microwave pulse $r(t)$ (blue).}
	\label{fig:apparatus}
	\end{figure*}
To study the system frequency response to a \emph{direct excitation of the material} like that related to the searched particle in ferromagnetic haloscopes, we photoexcite the material with 1064\,nm-wavelength, 11\,ps-duration laser pulses and measure the power stored in the HS modes for several values of applied magnetic field. 
At 1064\,nm, the absorption coefficient of YIG is $\sim 10$\,cm$^{-1}$ \cite{Wemple:1974aa,Scott:1974aa,PhysRevB.48.16407}, corresponding to the transition $^6A_{1g}(^6S)\rightarrow ^4T_{1g}(^4G)$ between electronic levels in the octaedral crystal field configuration \cite{Wang:2018aa}. 
In addition, the laser beam is focused within the sphere and we can thus reasonably assume that the 0.1\,mJ energy of the pulse is almost entirely absorbed in the material. A fraction of this energy is transferred from the excited electrons to magnetic oscillations of the material, including the uniform precession mode used in the ferromagnetic haloscope.
The optical pulse corresponds to a broadband excitation on the HS, allowing to study the system dynamics by considering the stimulus as frequency-independent. Therefore this optical excitation  is well represented by the last term in  Eq.\,(\ref{eq:ham}), mimicking the axion interaction for the present purpose of demonstrating the spin magnetometer dynamical bandwidth. 

We accomplish the strong cavity regime by coupling the Kittel mode, i. e. the uniform precession ferromagnetic resonance\cite{Kittel2004, PhysRev.143.372}, to the TE102 mode of a rectangular cavity, whose resonance frequency is $\omega_c\simeq(2\pi)\,4.7\,$GHz and the linewidth is $\gamma_c\simeq(2\pi)\,1.1\,$MHz. This magnetic dipole coupling is strengthened when the magnetic field amplitude of the chosen cavity mode is maximum at the location of the spin ensemble.
By setting a static field $B_0$ such that $\omega_m=\gamma B_0 = \omega_c$, we measure the coupling coefficient $2g_{cm}\simeq57\,$MHz, and $\gamma_m$ through the linewidth of the hybrid mode $\gamma_h$ as $\gamma_m=2\gamma_h-\gamma_c\simeq3.5\,$MHz (see Fig.\,\ref{fig:apparatus}a).

The laser pulses deposit energy in the YIG sphere, which is electromagnetically transduced to cavity excitations.
The power deposited in the hybrid mode is dissipated in a time $1/\gamma_h=\tau_h$, and is measured through an antenna coupled to the cavity. 
Hence, when the infrared laser pulse excites the YIG sphere, the representative signal $r(t)$ shown in Fig.\,\ref{fig:apparatus}b is detected with the heterodyne microwave receiver detailed in the Methods section.
The amplitude of the down-converted $r(t)$ signal is comparable with the receiver noise when $B_0$ is such that the system is far from the anticrossing point, ruling out a direct cavity excitation.

\begin{figure}[h!]
\centering
\includegraphics[width=.4\textwidth]{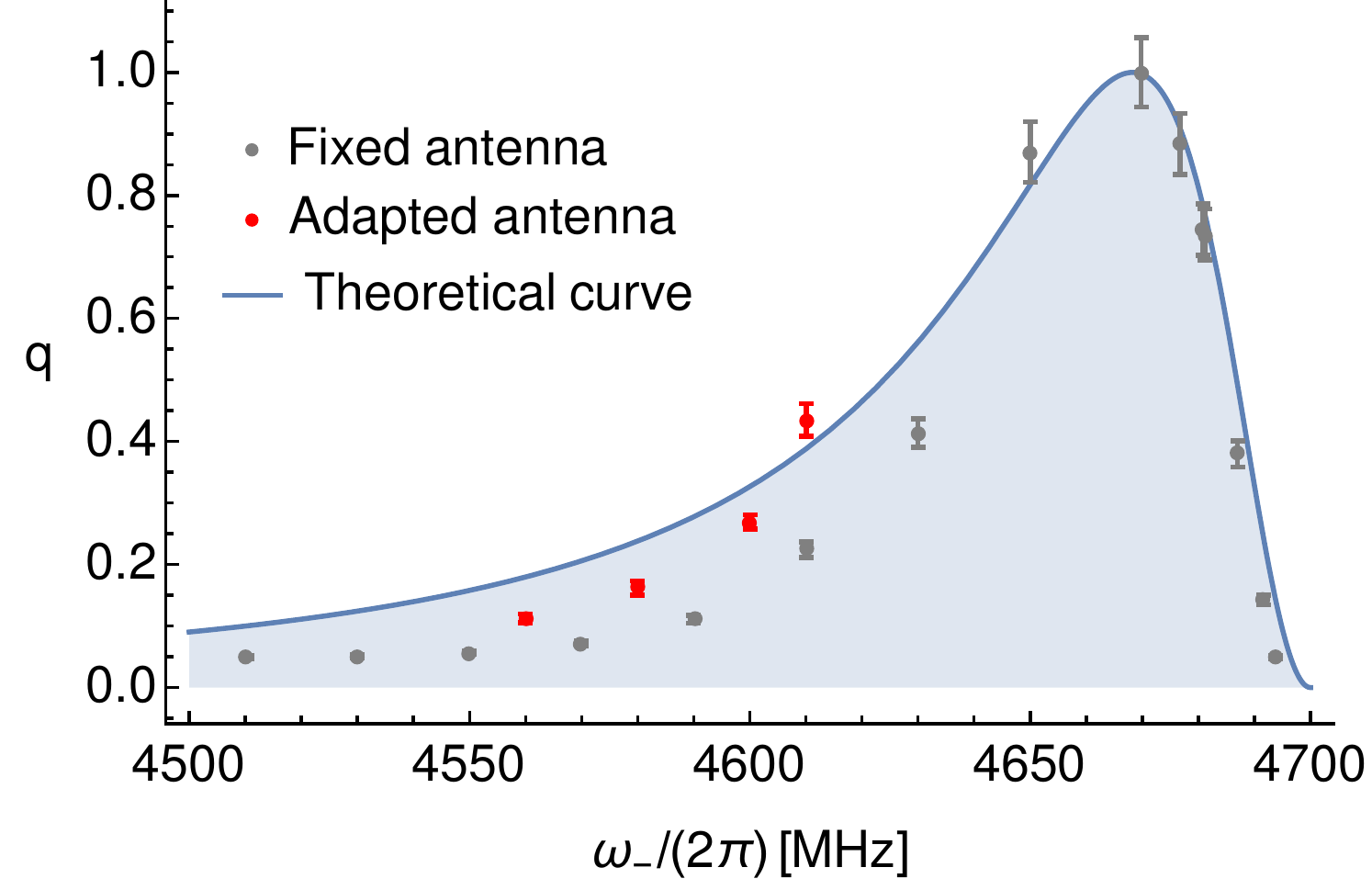}
\caption{Theoretical curve and experimental data. The blue curve has been calculated with the parameters reported in Fig.\,\ref{fig:bw}. The maximum measured signal was taken to be unitary and the other points were scaled consequently. The gray points were collected with a fixed antenna optimized for measurements around 4.67\,GHz, while for the red ones its coupling was, to some extent, adapted for the respective frequency. The data points and relative errorbars are obtained as mean and standard deviation of repeated measurements.}
\label{fig:res}
\end{figure}

To obtain the transduction coefficient from magnons to photons realised by the strong coupling, we record $r(t)$ for several values of the hybrid frequency $\omega_-$, which is varied through the $B_0$ field.
The experimental results are compared to the theoretical prediction in Fig.\,\ref{fig:res}. The transduction curve $q(\omega_-)=P_{ac}(\omega_-)/\max (P_{ac}(\omega_-))$ has been derived from Eq.\,(\ref{eq:p_bw}), and is the normalised top-left projection of the dispersion plot reported in Fig.\,\ref{fig:bw}.

Each measured value is instead obtained with the integral:
\begin{equation}
q_r(\omega_-)=\int_{t_0}^{t_\mathrm{laser}+4\tau_h}|r(t)|^2 \mathrm{d}t,
\label{eq:3_laserdatapoint}
\end{equation}
where $t_0$ is given by the laser pulse (orange curve in Fig.\,\ref{fig:apparatus}b). 
As the laser system exhibits shot-to-shot intensity fluctuation of the output pulses, for each $\omega_-$ the mean and standard deviation of the data is obtained by averaging hundreds of $q_r(\omega_-)$.
In Fig.\,\ref{fig:res} data points are plotted normalised with respect to the one with larger amplitude at 4.67\,GHz.
It is important to note that no additional elaboration has been carried out on the data, nor fitting procedure on the curve reported in Fig.\,\ref{fig:res}. They indeed display a good agreement for $\omega_->4.65\,$GHz, with a slight discrepancy at lower frequencies that can be explained in terms of mismatched transmission line. In fact, in the model we implicitly assume a condition of critical coupling (i.e. the extracted power is maximum), whereas the coupling between the receiver antenna and the cavity mode changes for different values of $\omega_{-}$. 
When the coupling is optimised (red data points in Fig.\,\ref{fig:res}), the agreement between experiment and theory significantly improves.
The system, see Fig.\,\ref{fig:bw}, should behave symmetrically for the resonance $\omega_+$.

\section*{Discussion}
Devising a simple and yet predictive model of a HS-based transducer is a key ingredient to understand the behaviour of a ferromagnetic haloscope. 
We used a simple second-quantisation model based on a system of two strongly coupled oscillators, a microwave cavity mode and a magnetostatic mode, to describe the transduction of pure magnetic excitations to microwave photons. 
The modelled physical system includes an external power injection through the magnetostatic mode.
This magnetisation oscillation can be given e.\,g. by Dark Matter axions, which, interacting coherently with magnons, would resonantly deposit power in the HS. The transducer converts magnons into photons which can be collected by an antenna. 
From this model we derive the HS transduction coefficient as a function of the external magnetic field, an easily tunable parameter which changes the resonant frequencies of the HS, and thus the transduction frequency.

To validate the model and to measure the dynamical bandwidth of a photon-magnon HS transducer, we introduce a new optical method to realise a selective excitation of the HS magnetic component, as is the case for axion-DM interactions in the haloscope.
This is fundamentally different from the calibration procedure used in Primakoff haloscopes\cite{PhysRevLett.120.151301,PhysRevLett.124.101303}, wherein rf power injection is accomplished through an antenna coupled to the cavity mode.
The axion-field  sensitivity of this setup is not discussed as the apparatus was not optimized to this aim. The sensitivity of a similar apparatus, and the one of an optimized haloscope, are reported in Ref.s \onlinecite{BARBIERI1989357} and \onlinecite{quaxprl}, respectively. We note that the magnetic field tuning of the apparatus can be combined with the variation of the cavity resonant frequency\cite{PhysRevLett.124.101303,PhysRevLett.120.151301,doi:10.1063/5.0015660} to further increase the total bandwidth of the apparatus.

In a ferromagnetic haloscope, a spectroscopic characterisation provides the HS parameters, namely its resonant frequencies, linewidths and couplings, that we insert in our analytical model.
By varying the static magnetic field, we change the transduction frequency and experimentally reconstruct the dynamical bandwidth, that is found to be in agreement with the function obtained by the model. 
Our findings represent the first confirmation of the wide frequency tunability of ferromagnetic haloscopes. Axion masses can in principle be probed by an optimised haloscope in a range up to a few GHz, provided that large bandwidth quantum-limited amplifiers, like travelling-wave Josephson parametric amplifiers\cite{Macklin307,PhysRevX.10.021021}, are employed as a first stage of amplification.

The present model can thus be extended to describe more sophisticated ferromagnetic haloscope configurations, namely including ten YIG spheres, as reported in Ref.\,\onlinecite{quaxprl}, where the transduction efficiency has been used to calculate the upper limit on the axion-electron coupling constant. 
Whilst we envision the upgrade of the ferromagnetic haloscope to even larger scales to boost the axion signal, we are aware of limitations that might arise in the HS due to the consequent large hybridization and the corresponding broader dynamical bandwidth. The use of larger spheres\cite{tobar}, multiple samples\cite{quaxepjc} or both\cite{quaxprl}, leads to higher order modes, size-effects or multi-spheres disturbances that are not taken into account in the modelled two oscillator HS. A consistent description of such complex apparatus can be obtained by adding additional modes in our model, that can be easily extended to an arbitrary number of coupled oscillators.

These optical tests will be fundamental to get end-to-end calibration in precision measurement apparatuses as those previously mentioned, provided the conversion factor of the underlying process is precisely known. For instance, a cw laser whose intensity is modulated at the relevant frequency might be used to mimic the effective AC magnetic field via the inverse Faraday effect\cite{PhysRevB.93.174427}.
Moreover, similar phenomena are studied for optical manipulation of magnons\cite{PhysRevLett.118.107205,PhysRevB.93.174427}, hence setups like the one presented in this work could be useful to understand the conversion capabilities of HSs.

The study of HSs is continuously growing and cross-fertilise multiple physical topics.  In the last years non-Hermitian physics phenomena are tackled by coupling photons and magnons, and exceptional points and surfaces were experimentally demonstrated\cite{PhysRevLett.124.053602,Zhang2017,PhysRevLett.123.237202,doi:10.1063/1.5144202,PhysRevB.99.214415}. In general, our scheme allows for a selective power injection in HSs, feature which can be used to study them on a fundamental level\cite{Wolz2020}.

\section*{Methods}
\subsection*{Ultrafast laser}
The laser system in Fig.\,\ref{fig:heterodyne} consists of a passively mode-locked oscillator, followed by a quasi-CW, two-stage Nd:YVO$_4$ slabs amplifier. 
From a stable train of $8$ ps-duration, 1064\,nm-wavelength pulses at 60\,MHz repetition rate, an extra-cavity acousto-optic (AO) pulse-picker selects, with adjustable repetition rate, nJ energy pulses that are amplified up to an energy of $100\,\mu$J. 
The amplifier slabs are pumped by a 150\,W peak power quasi-CW laser diode array, synchronized with the oscillator pulses sampled by the AO pulse picker. This has been accomplished by clocking the laser electronics with a secondary output beam of the oscillator by means of a photodiode. The amplified pulses duration is $11$ ps, and their repetition rate is 30\,Hz.

\subsection*{Hybrid system}
The magnetic material is a 2\,mm diameter spherical YIG single crystal. The sphere is glued to a ceramic rod to suspend it in its position, with the easy axis aligned to the static magnetic field. The $B_0$ field is supplied by a small electromagnet. The cavity is a rectangular copper cavity, whose dimensions are 98\,mm$\times$12.6\,mm$\times$42.5mm. The TE102 mode resonates at 4.7\,GHz and the quality factor is 4300, estimated with an $S_{21}$ measurement. The sphere is located at the center of the cavity, where the rf magnetic field is maximum, and perpendicular to the direction of $B_0$.

\subsection*{Readout and heterodyne}
\begin{figure}
	\centering
	 \includegraphics[width=.45\textwidth]{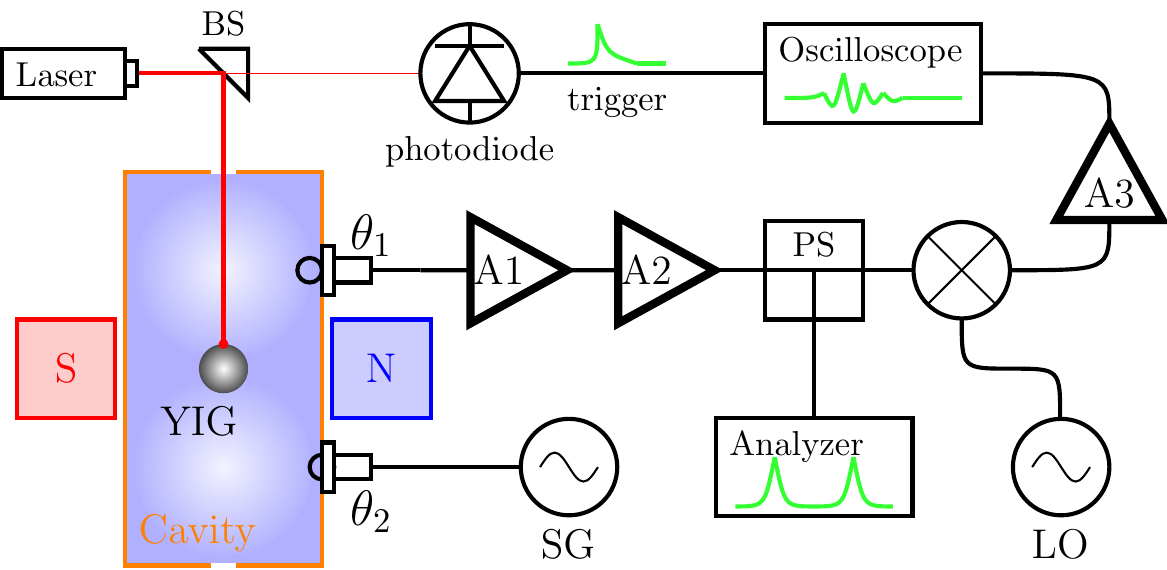}
	\caption{Scheme of the experimental setup. The 2\,mm-diameter YIG sphere is hosted inside a 98\,mm$\times$12.6\,mm$\times$42.5\,mm rectangular microwave cavity resonator, at the position of maximum magnetic field intensity for the TE102 mode (as indicated by the color map filling the cavity) to allow for maximising the coupling between cavity photons and magnons. Picosecond-duration, 1064\,nm-wavelength laser pulses are used to excite the magnetostatic modes in the YIG sphere, and the hybrid system dynamics is investigated through a heterodyne receiver coupled to the cavity mode. The signal is amplified before mixing with the output of a local oscillator (LO), further amplified and recorded at the oscilloscope for several values of the external magnetic field. A signal generator (SG), which is weakly coupled through an inductive loop to the cavity mode, combined with the power splitter (PS) and the spectrum analyser are used to obtain cavity and HS transmission spectra displayed in Fig.\,\ref{fig:apparatus}b. See text for further details.}
	\label{fig:heterodyne}
	\end{figure}

Two antennas, $\theta_1$ and $\theta_2$, are coupled to the rectangular cavity as sketched in Fig.\,\ref{fig:heterodyne}. The coupling of $\theta_2$, connected to the microwave oscillator SG, is fixed, while the coupling of $\theta_1$ can be varied by changing its position. The weakly coupled $\theta_2$ is used to inject microwave power, and thus perform spectroscopic measurements of the HS.
The position of $\theta_1$ is chosen by doubling the hybrid uncoupled linewidth when $\omega_c=\omega_m$, so that the coupling is close to critical,  and remains the same within the measurement.
The signal collected by $\theta_1$ is amplified by A1 and A2, two HEMT (high electron mobility field effect transistors) low-noise amplifiers. It is then split between a spectrum analyser, useful to acquire the transmission measurements of the HS, and a heterodyne that down-converts the signal to lower frequencies using a mixer and a local oscillator (LO).
The down-converted signal, whose band is shown in green in Fig.\,\ref{fig:apparatus}a, is further amplified by A3 and acquired by an oscilloscope triggered by the signal of a photodiode, as shown in Fig.\,\ref{fig:apparatus}b. 
The frequency of the local oscillator is adjusted whenever the $B_0$ field is changed, in order to keep the hybrid mode frequency $\omega_-$ within in the down-converted band. The frequency of $r(t)$ in Fig.\,\ref{fig:apparatus}b is the difference between the ones of the local oscillator and of the hybrid mode.
The total gain of the amplification chain is of order 71\,dB, and the noise temperature of A1 is about 40\,K, thus the noise of the apparatus is essentially due to room temperature thermodynamical fluctuations.

\section*{Data availability}
The data collected during this study are available from the corresponding author upon reasonable request.

\newpage
\bibliography{hybridSystemDynamics}


\section*{Acknowledgements}
The authors would like to thank Riccardo Barbieri for the help in the development of the theoretical model and Federico Pirzio for the advice on the laser system. We acknowledge the work of Enrico Berto, Mario Tessaro, and Fulvio Calaon, which contributed in the realization of the mechanics and electronics of the experiment.

\section*{Author contributions statement}
N.C. conceived the experiment and performed the measurements with the help of R.D., C.B. G.C. G.R. contributed to the construction of the setup, N.C. C.B. wrote the manuscript. All the author discussed the results and the manuscript.

\section*{Competing interests}
The authors declare no competing interests.

\section*{Additional information}
Correspondence and requests for materials shold be addressed to N.\,C. or R. D.

\end{document}